\documentclass[aps,prd,preprint,groupedaddress,showpacs,showkeys,onecolumn,10pt]{revtex4-1}
\usepackage{lipsum}
\usepackage{slashed} 
\usepackage{amsmath} 
\usepackage{tabularx}
\usepackage{color}
\usepackage{graphics}
\usepackage[dvips]{graphicx}
\usepackage{epsfig}

\usepackage[latin9]{inputenc}
\usepackage{booktabs}

\usepackage{amsthm}
\usepackage{amssymb}
\usepackage{graphicx}
\usepackage{lineno} % Include the package in the preamble

\usepackage{amsfonts}
\usepackage{newlfont}
\usepackage{color}

\usepackage{babel}
\usepackage{caption}
\usepackage{hyperref}

\makeatother

\begin{document}
\title{Production cross sections of charm-beauty mesons in proton-nucleus
and nucleus-nucleus collisions at LHC}
\author{Madeeha Nazish}\affiliation{Centre For High Energy Physics, Punjab
University, Lahore(54590), Pakistan. }
\email[]{diha.lifeline@gmail.com}

\author{Faisal Akram}
\email[]{faisal.chep@pu.edu.pk}
\affiliation{Centre For High Energy Physics, Punjab University,
Lahore(54590), Pakistan.}

\begin{abstract}
\noindent In this work, we calculate the production cross sections of
the 1S and 1P states of the $B_{c}$ meson for proton-nucleus and nucleus-nucleus
colliding beam experiments at LHC and RHIC. We provide estimates
for the Au nucleus at RHIC energies and for the Pb and Xe nuclei at LHC energies.
Previous estimates of these cross sections have ignored quark anti-quark
annihilation diagrams and are confined to S states for Pb and Au nuclei.
The cross sections of the P states are also equally important as these
states lie below the BD threshold and cascade into $B_{c}$ ground states
with 100\% probability. In addition to calculating the $p_{T}$, rapidity,
and pseudo-rapidity distributions of the cross sections, we also determine
the nuclear modification factors $R_{pA}$ and $R_{AA}$ for $A=\textrm{Pb}$,
$\textrm{Xe}$, and $\textrm{Au}$. The results provide a baseline
for the upcoming experimental studies of $B_{c}$ mesons in pp, pA, and
AA collisions in RUN III of LHC and could help develop experimental
strategies for identification of hitherto undiscovered states of the $B_{c}$
meson.
\end{abstract}

\pacs{}

\keywords{}

\maketitle

\section{Introduction}

The charm-beauty meson $(B_{c})$ is a mixed bound state of heavy
charm and bottom quarks. It was first discovered in 1998 by the CDF Collaboration
at Tevatron (Fermilab, USA) through its semi-leptonic decay $B_{c}\rightarrow J/\psi lv_{l}$
\citep{CDF:1998tqc}. CDF measured its mass as $6.40\pm0.39\pm0.13$
GeV and its lifetime as $0.46_{-0.16}^{+0.18}\pm0.03$ ps. Later,
it was produced by the D0 collaboration \citep{D0:2007gzu} at Tevatron
and LHCb experiment \citep{LHCb:2012ag} at LHC (Large Hadron Collider)
in CERN. Its current averaged mass and lifetime, as reported in PDG
(Particle Data Group) \citep{ParticleDataGroup:2024cfk}, are $6.275\pm0.8$
GeV and $0.510\pm0.009$ ps respectively. Recently, the 2S state of
the $B_{c}$ meson was discovered by the ATLAS experiment in LHC by
its strong decay $B_{c}(2S)\rightarrow J/\psi\pi$ \citep{ATLAS:2014lga}.
Its measured mass was $6.842\pm4\pm5$ GeV. This discovery was later
confirmed by the CMS \citep{CMS:2019uhm} and LHCb \citep{LHCb:2019bem}
experiments. Nevertheless, further measurements of the lifetime and
branching ratios of its decay modes are still required, although some
decay events have been observed. 

Being the only heavy quark meson carrying mixed quark flavors, it
provides unique conditions to study the dynamics of the heavy quarks.
Unlike other heavy quark mesons (charmonia and bottomonia), charm-beauty
mesons, below the BD threshold ($\approx7.144$ GeV), cannot disintegrate
through strong decays \citep{Godfrey:2004ya}. Consequently, they
are more stable and relatively easy to study. There are at least two
to three S states, two P states, one D state, and possibly one F state
of $B_{c}$ lying below the threshold of opening strong decays \citep{Godfrey:2004ya,Asghar:2019qjl}.
These excited states cascade through radiative and hadronic transitions
to the $B_{c}$ ground state $(1^{1}S_{0})$, which finally decays through
the weak interaction. The decay chains of the excited $B_{c}$ provide
us with unique signatures for their identification and reconstruction in
the experimental studies. The cleanest and most widely used decay
channel to detect the $B_{c}$ ground state is $B_{c}^{\pm}\rightarrow J/\psi\pi^{\pm}\rightarrow(\mu^{+}\mu^{-})\pi^{\pm}$,
having a combined branching ratio of 0.013\% and a detection efficiency of
about 2\% \citep{Godfrey:2004ya,LHCb:2024nlg}. The $B_{c}(1^{3}S_{1})$
decays into $B_{c}(1^{1}S_{0})\gamma$ through the electromagnetic transition.
However, the emitted soft photon of energy $E_{\gamma}\sim60$ MeV
escapes detection. This issue is not expected to be resolved at LHC
RUN III, so direct detection of $B_{c}(1^{3}S_{1})$ would remain a challeng;
instead, it gives a feed-down contribution to the $B_{c}$ ground state.
The spin singlet $(1^{1}P_{1})$ and triplet $(1^{3}P_{J=0,1,2})$
states decay to $B_{c}(1^{1}S_{0})\gamma$ and $B_{c}(1^{3}S_{1})\gamma$,
respectively. Since the physical states $1P_{1}$ and $1P'_{1}$ are mixed
with the $1^{1}P_{1}$ and $1^{3}P_{1}$ states, both can decay into
$1^{1}S_{0}$ and $1^{3}S_{1}$ states. Thus, four 1P states cascade
to the $B_{c}$ ground state through six decay chains. The energy of the
emitted photon in P to S transition is expected to be $\sim400$
MeV and can easily be resolved in LHC. However, the soft photon produced
through subsequent decay of $B_{c}(1^{3}S_{1})$ remains undetectable.
So, if we reconstruct 1P states through the invariant mass of $B_{c}(1^{1}S_{0})\gamma$,
we expect six overlapping peaks in which four are displaced due to
missing soft photon energy. Therefore, to accurately map the peaks
to the corresponding six decay chains, we are required to estimate
the corresponding number of events using the knowledge of the production
cross sections of the 1P states. Recently, LHCb has observed 1P states
as a broad structure peak using combined data of RUN I and II corresponding
to integral luminosity of 9 $\textrm{fb}^{-1}$ \citep{LHCb:2025uce}.
RUN III is expected to give additional data of 20-25 $\textrm{fb}^{-1}$
for LHCb, and with slightly improved energy resolution, we expect
that it would help in resolving peaks corresponding to the 1P states
and providing accurate measurements of their production cross sections.
Generally, RUN III's higher luminosity and improved detector performance
across the board would significantly enhance the visibility of the 1P
and other excited states below the BD threshold. For a comprehensive review
on search strategies for the excited states of the $B_{c}$ meson, we
refer to Refs. \citep{Asghar:2019qjl,Godfrey:2004ya}. It is, therefore,
crucial to calculate the production cross sections of the excited states of the $B_{c}$ meson to provide the estimates for the corresponding numbers
of cascade events. 

In this work, we calculate the production cross sections of the $B_{c}$ states
in pp (proton-proton), pA (proton-nucleus), and AA (nucleus-nucleus)
collisions at the LHC and RHIC energies. The $B_{c}$ production is particularly
interesting in pA and AA collisions, where several other effects may
significantly alter the estimates obtained from the hard production
mechanism. The most widely anticipated effect of quark-gluon plasma
(QGP), already very well studied for other heavy quark systems including
charmonia \citep{Grandchamp:2003uw,Arleo:2012rs,Fujii:2013yja,Ma:2015sia,Ducloue:2015gfa,Matsui:1986dk}
and bottomonia \citep{Grandchamp:2005yw,Vogt:2010aa}, is the suppression
of heavy quarkonia. The color Debye screening during the QGP phase of
the heavy-ion collisions can dissociate heavy quarkonia resulting
in anomalous suppression in the observed production rates. This mechanism
also predicts a pattern of sequential suppression in which the states
of higher radii are affected before the smaller one. Remarkably, CMS
collaboration has recently measured this sequential suppression in
radially excited states of $\Upsilon$ \citep{CMS:2012gvv,ALICE:2020wwx}.
Owing to the unique nature of this effect, the observation could be
regarded as robust evidence of QGP formation in LHC. Other than
the effect of QGP, the observed yield of $B_{c}$ depends on nuclear
shadowing effect \citep{EuropeanMuon:1987obv,Gavin:1990gm}, energy
loss of partons while passing through nuclear matter before involving
in the hard process, dissociation of heavy-quark pairs $Q\bar{Q}$ while
passing through nuclear matter, possible regeneration effect \citep{Liu:2012tn},
and dissociation through inelastic interaction with light mesons in
hadronic phase of heavy-ion collision \citep{Irfan:2015qxa}.
Unlike pp collision, the production of $B_{c}^{+}$ can proceed through
the combination of charm ($c$) and anti-bottom ($\bar{b}$) quarks
produced in different nucleon-nucleon interactions of participation
nucleons of colliding heavy-nuclei \citep{Liu:2012tn}. Moreover, this
effect may also be boosted by the formation of QGP producing many
unpaired charm and bottom quarks due to color Debye screening \citep{Letessier:2002ab,Alberico:2013pha,Liu:2012tn}.
If $B_{c}$ production is significantly enhanced by the regeneration
mechanism so that the total number of events is much larger than what
we can possibly produce from pp collisions, then heavy-ion collision
experiments can provide a new arena to complete the unfinished task
of $B_{c}$ spectroscopy in the coming days. As mentioned earlier,
the observed $B_{c}$ production may also be affected by its dissociation
via inelastic interactions with comoving light mesons ($\pi$, $\rho$,
etc.) in the hadronic phase. The exact amount of dissociation depends
on the $B_{c}$ hadronic cross sections that have been estimated in
several studies \citep{Lodhi:2011zz,Akram:2011zz}. A recent work
of Ref. \citep{Irfan:2015qxa} indicates that suppression caused by
this effect is small but not negligible.

A first step towards a thorough study of the $B_{c}$ meson in heavy-ion
collisions is to obtain estimates of its inclusive production cross
sections. Theoretical studies of these production cross sections for
hadronic colliders are very well-established using fragmentation \citep{Cheung:1993pk,Cheung:1993qi,Cheung:1995ir,Cheung:1995ye}
and NRQCD approaches \citep{Chang:1992jb,Chang:1994aw,Chang:1996jt,Chang:2003cr,Chang:2004bh};
both using the QCD factorization theorem. However, the study of
$B_{c}$ production cross sections in relativistic heavy-ion collisions
is still in the initial stages. Present estimates of $B_{c}$ production
cross sections in pA and AA collisions lack the inclusion of quark
anti-quark annihilation diagrams and calculations are confined to
1S states \citep{Chen:2018obq}. The knowledge of the cross sections
of the 1P states is equally important because they lie below the BD threshold
and cascade to the $B_{c}$ ground state with 100\% probability. In
Ref. \citep{Chen:2018obq}, the cross sections of 1S states are calculated
at RHIC and LHC energies for Au and Pb targets, respectively. Since
LHC has started collecting the data for p-Xe and Xe-Xe colliding beams
as well, extending the theoretical estimates to Xe target is also
important.

In this work, we reproduce and extend the work of Ref. \citep{Chen:2018obq}.
We calculate the inclusive production cross section of both S and
P states of $B_{c}$ meson for pp, pA and AA collisions at LHC and
RHIC energies. At LHC energy, we calculate the cross sections for
p-Pb, p-Xe, Pb-Pb, Xe-Xe collisions, and at RHIC energy we consider
p-Au and Au-Au collisions. For pA and AA collisions, we use nuclear
parton distribution functions, which include the effect of nuclear
shadowing. To quantify the effect of nuclear shadowing in
pA and AA collisions, we calculate the $p_{T}$ and rapidity distributions
of the nuclear modification factors of 1S and 1P states of $B_{c}$ meson
for pA and AA collisions. We anticipate that these results would provide
a baseline for upcoming experimental studies of $B_{c}$ in pp, pA,
and AA collisions in RUN III of LHC.

The paper is organized as follows. In Sec. 2, we cover the details
of the methodology used to calculate the cross sections. In Sec. 3,
the values of the parameters used in computing the cross sections
are discussed. In Sec. IV, we present the numerical results of cross
section along with uncertainties, distributions of cross sections
and nuclear modification factors. In Sec. V, we discuss the results and
give concluding remarks.

\section{Methodology}

The most likely mechanism of producing a $B_{c}^{+}$ meson is via a hard
process in which two pairs $b\bar{b}$ and $c\bar{c}$ are produced
through interactions between the partons of the colliding hadrons.
A $B_{c}^{+}$ meson emerges in the process when two collinear $\bar{b}$
and $c$ quarks combine. By employing NRQCD factorization approach \citep{Caswell:1985ui,Thacker:1990bm,Bodwin:1994jh}
along with the QCD factorization theorem, we can write the production
cross section of a $B_{c}^{+}$ state as follows:
\begin{equation}
d\sigma_{(AB\rightarrow\mathcal{B}_{c}^{+}+X)}=N_{A}N_{B}\sum_{n,i,j}\int dx_{1}dx_{2}f_{i}^{A}(x_{1},\mu_{F})f_{j}^{B}(x_{2},\mu_{F})d\hat{\sigma}_{(ij\rightarrow c\bar{b}[n]X)}\left\langle \mathcal{O}_{n}^{B_{c}^{+}}\right\rangle ,\label{1}
\end{equation}

\noindent where $d\hat{\sigma}_{(ij\rightarrow c\bar{b}[n]X)}$ is
the perturbatively calculated partial cross section of producing $c\bar{b}$
pair in state with quantum numbers $n\equiv^{2S+1}L_{J}^{(c)}$, where
$S$, $L$, and $J$ are spin, orbital and total angular momentum,
respectively, and $c=1,8$ for color singlet and octet states, $\left\langle \mathcal{O}_{n}^{B_{c}^{+}}\right\rangle $
are long-distance-matrix elements (LDMEs) that describe the transition
of perturbative $c\bar{b}[n]$ states to non-perturbative physical
state of $B_{c}^{+}$ meson. In principle, the sum over $n$ encompasses
all possible perturbative states $c\bar{b}[n]$, requiring an infinite
number of LDME's $\left\langle \mathcal{O}_{n}^{B_{c}^{+}}\right\rangle $
and corresponding short distance coefficients $d\hat{\sigma}_{(ij\rightarrow c\bar{b}[n]X)}$.
However, velocity scaling rules \citep{Lepage:1992tx} of NRQCD imply
that only a finite number of contributions are required to get a result
accurate up to a certain order in $v$ and $\alpha_{s}$, where $v$
is the relative velocity of quark anti-quark pair. For S physical
states of $B_{c}^{+}$ only one $c\bar{b}[n]$ state contributes at
the leading order in $v$ in which $n$ matches the quantum numbers
of the physical state. Whereas, for P physical states two $c\bar{b}[n]$
states contribute at the leading order in $v$: one in which $n$
matches the quantum number and the other is a color octet state $^{1}S_{0}^{(8)}$
or $^{3}S_{1}^{(8)}$ for $^{1}P_{1}^{(1)}$ and $^{3}P_{J}^{(1)}$
physical states respectively. The functions $f_{i}^{A}(x_{1},\mu_{F})$
and $f_{j}^{B}(x_{2},\mu_{F})$ are parton distribution functions
(PDFs) of $i$ and $j$ partons in colliding particles $A$ and $B$,
where $A,B=p,Au,Pb,Xe$. For pp collisions $N_{A}=N_{B}=1$, for pA
collisions $N_{B}=1$ and $N_{A}$ is the mass number of the nucleus $A$ ,
and for AA collisions $N_{A}$ and $N_{B}$ are equal to mass numbers
of $A$ and $B$ nuclei, respectively. It is noted that Eq. \ref{1}
gives minimum-bias (MB) value of the cross section. In case of pA
and AA collisions we use nuclear PDFs incorporating the nuclear shadowing
effect. We use a separate set of nCTEQ15 LO PDF \citep{Kovarik:2015cma}
for Au, Pb, and Xe nuclei, respectively. For pp collisions, we use CTEQ6L1 PDF set
\citep{Pumplin:2002vw}. The sum over $i$ and $j$ corresponds to the
sum of all possible subprocesses in which $b\bar{b}$ and $c\bar{c}$
pairs are produced. However, at LO in $\alpha_{s}$, only two
subprocesses contribute.
\begin{eqnarray}
g+g & \rightarrow & c\bar{b}[n]+\bar{c}+b\\
q+\bar{q} & \rightarrow & c\bar{b}[n]+\bar{c}+b
\end{eqnarray}
Corresponding to the gluon-gluon fusion process, we have 36 Feynman diagrams
at leading order in $\alpha_{s}$ and 7 for the quark anti-quark annihilation
process. In this work, we have studied both subprocesses, though the dominant
contribution comes from the gluon-gluon fusion process. At next-to-leading
order (NLO), the Feynman diagrams enormously increase, making the computation
extremely difficult. A study at NLO is under preparation by the Authors.

\begin{figure}
\begin{centering}
\includegraphics[scale=0.6]{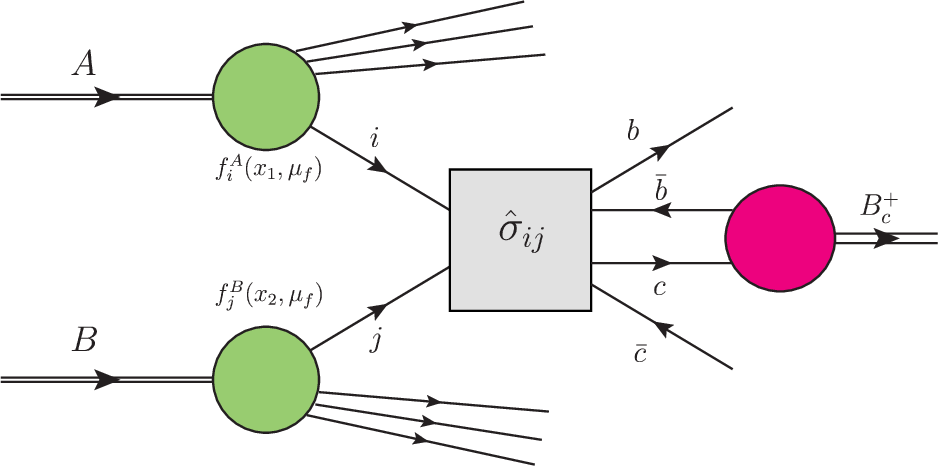}
\par\end{centering}
\caption{Diagramatic representation of production of $B_{c}$ meson through
partonic interaction $ij\rightarrow B_{c}X$ in a heavy-ion collision.\label{fig:AA-Bc}}
\end{figure}

\section{Input parameters}

At LHC, the collision energy for the proton-proton (pp) mode is $\sqrt{S_{pp}}=13$
TeV, with each proton beam having an energy of 6.5 TeV. In the pA
mode, the beam energy of the nucleus per nucleon is related to the
proton energy by $E_{N}=\frac{Z}{N_{A}}E_{p}$. In case of AA mode,
we use design values. The collision energies and corresponding beam
energies at LHC and RHIC are summarized in Table \ref{tab:cm-energies}.

\begin{table}
\centering{}\centering%
\begin{tabular}{ccc}
\hline 
Colliding Beams  & Collision Energy $\sqrt{S_{AB}}$(TeV)  & Beam energies $E_{A/B}$ (TeV) \tabularnewline
\hline 
$pp$  & 13  & $E_{p}=6.5$ \tabularnewline
$p$Pb  & 8.16  & $E_{p}=6.5,E_{\text{Pb}}=2.56$ \tabularnewline
PbPb  & 5.02  & $E_{Pb}=2.51$ \tabularnewline
$p$Xe  & 8.3  & $E_{p}=6.5,E_{\text{Xe}}=2.68$ \tabularnewline
XeXe  & 5.44  & $E_{\text{Xe}}=2.72$ \tabularnewline
$p$Au  & 0.2  & $E_{p}=0.150,E_{\text{Au}}=$0.07 \tabularnewline
AuAu  & 0.2  & $E_{\text{Au}}=0.1$ \tabularnewline
\hline 
\end{tabular}\caption{Collision energies in cm frame and corresponding beam energies for
different colliding modes. In case of pA mode, the energy of the nucleus
per nucleon is fixed by $E_{A}=E_{p}Z/N_{A}$, whereas in case of
AA mode, design value is used.\label{tab:cm-energies}}
\label{table1} 
\end{table}
Most of the other input parameters are taken same as in Ref. \citep{Chen:2018obq}.
We take heavy quark masses $m_{c}=1.5$ GeV and $m_{b}=4.9$ GeV.
The mass of $B_{c}$ is approximated by $m_{c}+m_{b}$. The renormalization
and factorization scales are taken same and set to $M_{t}=\sqrt{p_{T}^{2}+M_{B_{c}}^{2}}$,
where $p_{T}$ is the transverse momentum of $B_{c}$. The strong coupling
constant $\alpha_{s}$ is taken running. For isolated protons, we use
CTEQ6L1 PDF set \citep{Pumplin:2002vw}, whereas for Au, Pb, and Xe
nuclei, we use nuclear PDF sets nCTEQ15LO \citep{Kovarik:2015cma}.
To calculate the production cross-sections of S and P states at the leading-order
in $v$, we require the values of the following LDMEs:
\begin{eqnarray}
\left\langle \mathcal{O}^{B_{c}^{+}}[^{1}S_{0}^{(1)}]\right\rangle  & = & \left\langle \mathcal{O}^{B_{c}^{+}}[^{3}S_{1}^{(1)}]\right\rangle =\frac{\left\vert R(0)\right\vert ^{2}}{4\pi},\\
\left\langle \mathcal{O}^{B_{c}^{+}}[^{1}P_{1}^{(1)}]\right\rangle  & = & \left\langle \mathcal{O}^{B_{c}^{+}}[^{3}P_{0}^{(1)}]\right\rangle =\left\langle \mathcal{O}^{B_{c}^{+}}[^{3}P_{1}^{(1)}]\right\rangle =\left\langle \mathcal{O}^{B_{c}^{+}}[^{3}P_{2}^{(1)}]\right\rangle =\frac{3\left\vert R^{\prime}(0)\right\vert ^{2}}{4\pi},
\end{eqnarray}

\noindent where we have used heavy-quark spin symmetry to relate the
matrix elements of spin-singlet and triplet states. Heavy quark spin
symmetry is an approximate symmetry of NRQCD Lagrangian \citep{Bodwin:1994jh}
which hold at leading order in $v$. The radial wave function and
its derivative at origin can be obtained by applying non-relativistic
potential model. We also ignore small effects of hyperfine contact
interaction producing spin-splitting and spin-orbit tensor interaction
producing splitting in spin triplet P states. For 1S states, we take
radial wave function at origin $\left\vert R(0)\right\vert ^{2}=1.642$
GeV$^{3}$ and for 1P states $\left\vert R^{\prime}(0)\right\vert ^{2}=0.201$
GeV$^{5}$ \citep{Akbar:2018hiw,Eichten:1994gt,Eichten:1995ch}. Same
values for 1S states are also applied in Ref. \citep{Chen:2018obq}.
For color-octet matrix elements of S states, radial wave function
at origin cannot be obtained from non-relativistic potential model.
However, NRQCD velocity scaling rules show that these matrix elements
are one order higher in $v^{2}$ than the corresponding matrix elements
of singlet states. In Ref. \citep{Chang:2005bf}, this scaling property
is used to relate LDMEs of S states as following
\begin{equation}
\left\langle \mathcal{O}^{B_{c}^{+}}[S^{(8)}]\right\rangle \approx\Delta_{s}(v)^{2}\left\langle \mathcal{O}^{B_{c}^{+}}[S^{(1)}]\right\rangle ,
\end{equation}

\noindent where $\Delta_{s}(v)$ is of order $v^{2}$ and taken to
be 0.1. Same values of LDME's of octet states are also used in Ref.
\citep{Chen:2018obq}.

\section{Results}

\subsection{Cross sections}

\begin{figure}
\begin{centering}
\includegraphics[scale=0.9]{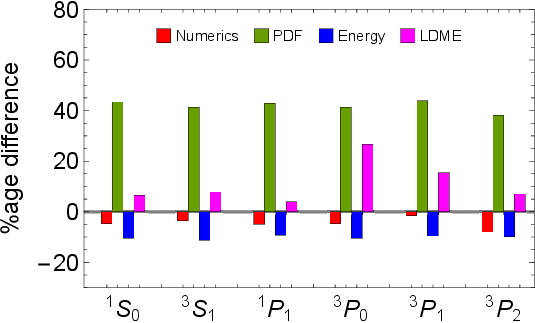}
\par\end{centering}
\caption{Breakdown of all systematic differences between our results and those
of Ref. \citep{Chang:2005bf}, in which CTEQ6L PDF set is used, at
$\sqrt{S}=14$ TeV, and $\left\vert R(0)\right\vert ^{2}=1.54$ GeV$^{3}$.
The difference due to numerical implementation is at most 5\%. \label{fig:pp_comp}}

\end{figure}

In table \ref{tab:pp}, we report our results of the production cross
sections of the $B_{c}$ meson for the 1S and 1P states in pp collision at LHC
energy. To validate the accuracy of our calculations, we also provide
a comparison with the results obtained in Ref. \citep{Chang:2005bf}
for the 1S and 1P states and \citep{Chen:2018obq} for the 1S states. The apparent
discrepancy between our results and that of Ref. \citep{Chang:2005bf}
is mainly attributed to the choice of a different PDF set and, to
a lesser extent, to slightly different values of LDME and c.m. energy,
as indicated in Fig. \ref{fig:pp_comp}, where we provide a detailed
breakdown of all systematic variations. It is not possible to identify
the exact reason for the quantitative difference with results of Ref.
\citep{Chen:2018obq}, as they do not specify the PDF set used and
whether $\alpha_{s}$ is taken running or constant. We also separately
provide the contribution to production cross sections from the gluon fusion
and $q\bar{q}$ annihilation processes in the table \ref{tab:pp}. These results
show that the contribution from $q\bar{q}$ annihilation process is less
than 1\% of the gluon fusion process for each state. In table \ref{tab:cross-pb},
we report our results of production cross sections of the 1S and 1P states
of the $B_{c}$ meson in pA and AA collisions for Pb nucleus at LHC energies
using nCTEQ15 LO PDF \citep{Kovarik:2015cma}. We also provide a comparison
with the results obtained in Ref. \citep{Chen:2018obq} for 1S states.
We use $m_{c}=1.5$ GeV, $m_{b}=4.9$ GeV, and $\mu_{F}=\sqrt{p_{T}^{2}+M_{B_{c}}^{2}}$,
and running $\alpha_{s}$, the same as used in calculating the production
cross section in the pp collision. The values of LDMEs estimated from
the wave-function and its derivative at origin are also same. In Ref.
\citep{Chen:2018obq}, authors used same PDF set and input parameters,
however, they do not specify whether $\alpha_{s}$ is taken running
or constant, so it not possible to identify the reason for apparent
quantitative difference. In table \ref{tab:cross-xe}, we report our
results of production cross sections of S and P states of the $B_{c}$
meson for Xe nucleus at LHC energy. At LHC energy, the contribution
of $q\bar{q}$ annihilation process is negligible, so all values reported
in tables \ref{tab:cross-pb} and \ref{tab:cross-xe} are for the gluon
fusion process. In table \ref{tab:cross-au}, we report our results
of production cross sections for Au nucleus at RHIC energy. A comparison
with the results obtained in Ref. \citep{Chen:2018obq} for S states
is also provided. At RHIC energy, it is expected that $B_{c}$ production
cross section may take a significant contribution from $q\bar{q}$ annihilation
process. In to order highlight it, we separately provide the contributions
to production cross sections from the gluon fusion and $q\bar{q}$ annihilation
processes for both p-Au and Au-Au collisions. These results show that
the contribution from $q\bar{q}$ annihilation process is about 10\%
of gluon fusion process for each state. All values of total cross
sections reported in tables \ref{tab:cross-pb}, \ref{tab:cross-xe},
and \ref{tab:cross-au} are obtained without any kinematical cuts on $p_{T}$
and $y$ (rapidity) of the $B_{c}$ meson. These calculations utilize  nCTEQ15 LO PDF sets,
which include error sets. The error sets are used to estimate the errors in the cross sections arising from PDF 
uncertainties, as reported
in these tables. Moreover, to study the impact of the choice of a particular
PDF set (nCTEQ15), we employe a selection of alternative nuclear PDF
sets for Pb and Xe nuclei at LHC energies. In table \ref{tab:alt_pdf_pb},
we compare the results of cross sections for p-Pb and Pb-Pb collisions
obtained using nCTEQ15 LO PDF sets with those employing EPPS16 \citep{Eskola:2016oht}
and nNNPDF3.0 \citep{AbdulKhalek:2019mzd} PDF sets. Similarly, in
table \ref{tab:alt_pdf_xe}, we compare the results of nCTEQ15 p-Xe
and Xe-Xe collisions results with nNNPDF3.0 set only, due to non-availability
of EPPS16 PDF sets for Xe nucleus. 

Leading-order results are usually very sensitive to the choice of
factorization scale and the masses of heavy quarks including $m_{c}$
and $m_{b}$. Given the mass of charm quark $m_{c}=1.5\pm0.1$ GeV,
we calculate the cross sections for three values of $m_{c}=1.4,$
$1.5$, and $1.6$ GeV, while taking $m_{b}=4.9$ GeV. The resulting
values of the cross sections for Pb, Xe, and Au nuclei are used
to determine uncertainties due to charm quark mass. Similarly, since the mass
of bottom quark $m_{b}=4.9\pm0.2$ GeV, we also calculate the cross
sections for three values of $m_{b}=4.7$, $4.9$, and $5.1$ GeV,
taking $m_{c}=1.5$ GeV. The resulting values
are used to determine errors due the bottom quark mass. In table \ref{tab:cross-err},
we combine all these results and report the values of the cross sections
with uncertainty due to heavy quark masses. First and second errors
correponds to $\Delta m_{c}=\pm0.1$ GeV and $\Delta m_{b}=\pm0.2$
GeV, respectively. It is noted that positive errors reported in the
table corresponds to $\Delta m_{c}=-0.1$ GeV and $\Delta m_{b}=-0.2$
GeV as the cross sections increase with decrease in heavy quark masses.
It is also found that a joint variation of charm and bottom quark
masses has a cumulative effect, i.e., the total uncertainty is almost
equal to the sum of uncertainties reported in the Table \ref{tab:cross-err}
due to independent variations in $m_{c}$ and $m_{b}$.

\begin{figure}
\begin{centering}
\includegraphics[scale=0.58]{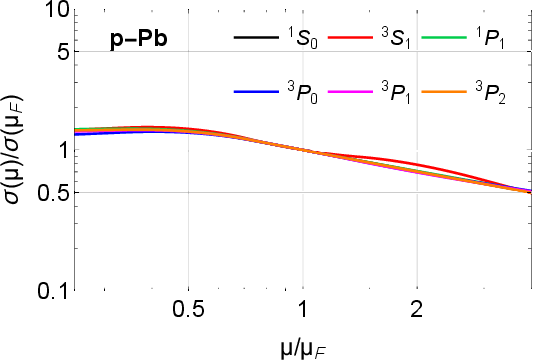}\includegraphics[scale=0.58]{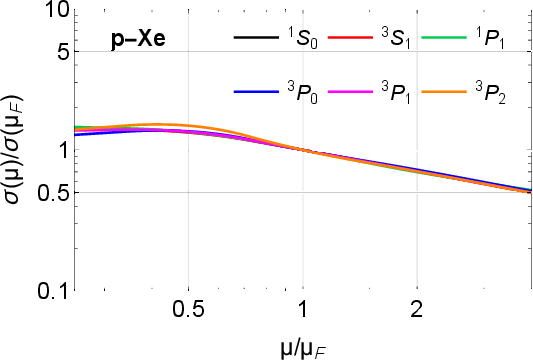}\includegraphics[scale=0.58]{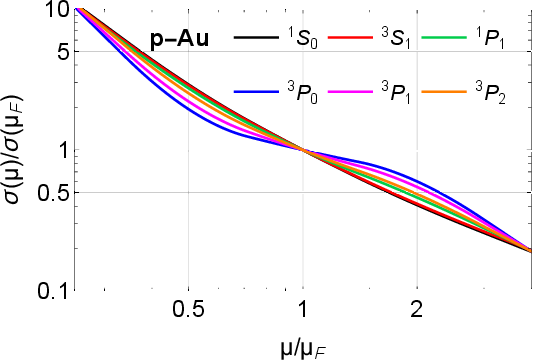}
\par\end{centering}
\caption{Variation of the production cross sections of the 1S and 1P states of the $B_{c}^{+}$
meson for p-Pb (left), p-Xe (middle), and p-Au (right) collisions
with factorization scale $\mu$, varied from $0.25\mu_{F}$ to $4\mu_{F}$.\label{fig:mu-pA}}
\end{figure}

\begin{figure}
\begin{centering}
\includegraphics[scale=0.58]{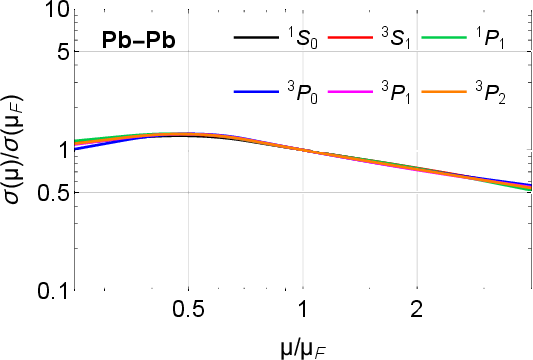}\includegraphics[scale=0.58]{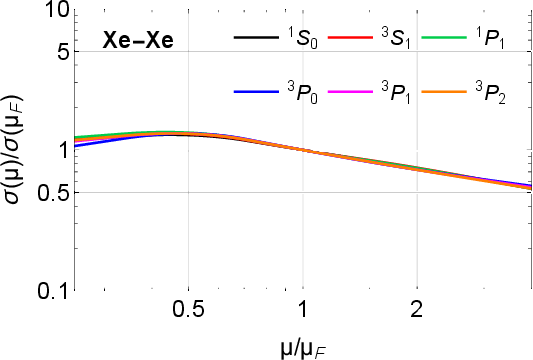}\includegraphics[scale=0.58]{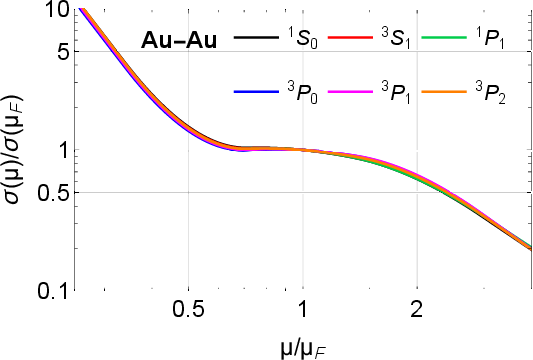}
\par\end{centering}
\caption{Variation of the production cross sections of the 1S and 1P states of the $B_{c}^{+}$
meson for Pb-Pb (left), Xe-Xe (middle), and Au-Au (right) collisions
with factorization scale $\mu$, varied from $0.25\mu_{F}$ to $4\mu_{F}$.\label{fig:mu-AA}}
\end{figure}

To study the sensitivity of estimates of cross sections to the
choice of factorization scale $\mu_{F}=\sqrt{p_{T}^{2}+M_{B_{c}}^{2}}$,
we vary it from $0.25\mu_{F}$ to $4\mu_{F}$. The results are given
in Figs. \ref{fig:mu-pA} and \ref{fig:mu-AA}. The plots show that
for all collision modes and states of the $B_{c}$ meson, the
variation is about $\pm25\%$ within the the range $0.25\mu_{F}$ to $2\mu_{F}$. However, for pAu and AuAu modes,
we observe relatively a large variation.

\begin{table}
\begin{centering}
\begin{tabular}{ccccc}
\hline
State & $\sigma(pp\rightarrow B_{c}^{+}X)$ nb & $\sigma(pp\rightarrow B_{c}^{+}X)$ nb & $\sigma(pp\rightarrow B_{c}^{+}X)$ nb \citep{Chang:2005bf} & $\sigma(pp\rightarrow B_{c}^{+}X)$ nb \citep{Chen:2018obq}\tabularnewline
\cmidrule{2-5} \cmidrule{3-5} \cmidrule{4-5} \cmidrule{5-5} 
 & (via gluon fusion) & (via $q\bar{q}$ annihilation) & (via gluon fusion) & (via gluon fusion)\tabularnewline
\hline
\hline 
$1^{1}S_{0}$ & $4.79\times10^{1}$ & $1.51\times10^{-1}$ & $3.56\times10^{1}$ & $4.24\times10^{1}$\tabularnewline
$1^{3}S_{1}$ & $11.9\times10^{1}$ & $9.32\times10^{-1}$ & $8.85\times10^{1}$ & $10.5\times10^{1}$\tabularnewline
$1^{1}P_{1}$ & $6.52\times10^{0}$ & $3.1\times10^{-2}$ & $4.92\times10^{0}$ & -\tabularnewline
$1^{3}P_{0}$ & $4.93\times10^{0}$ & $1.89\times10^{-2}$ & $3.23\times10^{0}$ & -\tabularnewline
$1^{3}P_{1}$ & $7.81\times10^{0}$ & $4.95\times10^{-2}$ & $5.27\times10^{0}$ & -\tabularnewline
$1^{3}P_{2}$ & $1.50\times10^{1}$ & $9.12\times10^{-2}$ & $1.18\times10^{1}$ & -\tabularnewline
\hline
\end{tabular}
\par\end{centering}
\caption{Production cross sections of the 1S and 1P states of the $B_{c}^{+}$ meson
at LHC energy $\sqrt{S}=13$ TeV. The second and third columns contain
our results for the cross sections via gluon fusion and $q\bar{q}$ annihilation,
respectively. The fourth and fifth columns contain the corresponding values taken
from Refs. \citep{Chang:2005bf} and \citep{Chen:2018obq}, respectively.\label{tab:pp}}
\end{table}

\begin{table}
\begin{centering}
{\small{}}%
\begin{tabular}{ccccc}
\hline 
{\small{}States} & {\footnotesize{}$\sigma(pPb\rightarrow B_{c}X)$ ($\mu\textrm{b}$)} & {\footnotesize{}$\sigma(PbPb\rightarrow B_{c}X)$ ($\mu\textrm{b}$)} & {\footnotesize{}$\sigma(pPb\rightarrow B_{c}X)$ ($\mu\textrm{b}$)
\citep{Chen:2018obq}} & {\footnotesize{}$\sigma(PbPb\rightarrow B_{c}X)$ ($\mu\textrm{b}$)
\citep{Chen:2018obq}}\tabularnewline
\cline{2-5} \cline{3-5} \cline{4-5} \cline{5-5} 
 & {\scriptsize{}($8.16$ TeV)} & {\scriptsize{}($5.04$ TeV)} & {\scriptsize{}($8.16$ TeV)} & {\scriptsize{}($5.04$ TeV)}\tabularnewline
\hline 
\hline 
{\small{}$1^{1}S_{0}$} & {\small{}$4.19\pm0.25$} & {\small{}$232\pm20$} & {\small{}$3.29$} & {\small{}$369$}\tabularnewline
{\small{}$1^{3}S_{1}$} & {\small{}$10.5\pm0.6$} & {\small{}$583\pm49$} & {\small{}$8.26$} & {\small{}$921$}\tabularnewline
{\small{}$1^{1}P_{1}$} & {\small{}$0.566\pm0.030$} & {\small{}$31.4\pm2.4$} & {\small{}-} & {\small{}-}\tabularnewline
{\small{}$1^{3}P_{0}$} & {\small{}$0.426\pm0.019$} & {\small{}$23.4\pm1.5$} & {\small{}-} & {\small{}-}\tabularnewline
{\small{}$1^{3}P_{1}$} & {\small{}$0.681\pm0.030$} & {\small{}$37.6\pm2.4$} & {\small{}-} & {\small{}-}\tabularnewline
{\small{}$1^{3}P_{2}$} & {\small{}$1.32\pm0.07$} & {\small{}$73.1\pm5$} & {\small{}-} & {\small{}-}\tabularnewline
\hline 
\end{tabular}{\small\par}
\par\end{centering}
\caption{{\small{}Production cross sections of the 1S and 1P states of the $B_{c}^{+}$
meson in p-Pb and Pb-Pb collisions at LHC energies. In second and
third columns, we provide our results using nCTEQ15 LO PDF sets.
In fourth and fifth columns, the estimates of Ref. \citep{Chen:2018obq}
are provided for the 1S states. The errors in our values are due to the
uncertainty in the PDF set.\label{tab:cross-pb}}}
\end{table}

\begin{table}
\begin{centering}
\begin{tabular}{ccc}
\hline 
{\footnotesize{}States} & {\footnotesize{}$\sigma(pXe\rightarrow B_{c}X)$ ($\mu\textrm{b}$)} & {\footnotesize{}$\sigma(XeXe\rightarrow B_{c}X)$ ($\mu\textrm{b}$)}\tabularnewline
\cline{2-3} \cline{3-3} 
 & {\footnotesize{}(8.30 TeV)} & {\footnotesize{}(5.44 TeV)}\tabularnewline
\hline 
\hline 
{\small{}$1^{1}S_{0}$} & $2.81\pm0.12$ & $104\pm6$\tabularnewline
{\small{}$1^{3}S_{1}$} & $7.01\pm0.29$ & $261\pm15$\tabularnewline
{\small{}$1^{1}P_{1}$} & $0.379\pm0.014$ & $14.1\pm0.7$\tabularnewline
{\small{}$1^{3}P_{0}$} & $0.285\pm0.009$ & $10.5\pm0.5$\tabularnewline
{\small{}$1^{3}P_{1}$} & $0.456\pm0.014$ & $16.9\pm0.7$\tabularnewline
{\small{}$1^{3}P_{2}$} & $0.882\pm0.030$ & $32.8\pm1.6$\tabularnewline
\hline 
\end{tabular}
\par\end{centering}
{\small{}\caption{{\small{}Production cross sections of the 1S and 1P states of the $B_{c}^{+}$
meson in p-Xe and Xe-Xe collisions at LHC energies using nCTEQ15
LO PDF. The errors in our values are due to the uncertainty
in the PDF set.\label{tab:cross-xe}}}
}{\small\par}
\end{table}

\begin{table}
\begin{centering}
\begin{tabular}{ccccccccc}
\hline 
States & \multicolumn{4}{c}{$\sigma(pAu\rightarrow B_{c}X)$ (nb)} & \multicolumn{4}{c}{$\sigma(AuAu\rightarrow B_{c}X)$ ($\mu\textrm{b}$)}\tabularnewline
\cline{2-9} \cline{3-9} \cline{4-9} \cline{5-9} \cline{6-9} \cline{7-9} \cline{8-9} \cline{9-9} 
 & ($gg$) & ($q\bar{q}$) & Total & Ref. \citep{Chen:2018obq} & ($gg$) & ($q\bar{q}$) & Total & Ref. \citep{Chen:2018obq}\tabularnewline
\hline 
\hline 
$1^{1}S_{0}$ & $8.03$ & $0.49$ & $8.52\pm0.50$ & $8.19$ & $1.11$ & $0.06$ & $1.17\pm0.09$ & $1.76$\tabularnewline
$1^{3}S_{1}$ & $18.86$ & $3.59$ & $22.5\pm1.1$ & $19.3$ & $2.60$ & $0.44$ & $3.04\pm0.22$ & $4.15$\tabularnewline
$1^{1}P_{1}$ & $1.06$ & $0.13$ & $1.19\pm0.06$ & - & $0.146$ & $0.017$ & $0.163\pm0.011$ & -\tabularnewline
$1^{3}P_{0}$ & $1.03$ & $0.08$ & $1.11\pm0.05$ & - & $0.143$ & $0.010$ & $0.153\pm0.008$ & -\tabularnewline
$1^{3}P_{1}$ & $1.38$ & $0.14$ & $1.52\pm0.08$ & - & $0.191$ & $0.018$ & $0.209\pm0.011$ & -\tabularnewline
$1^{3}P_{2}$ & $2.39$ & $0.33$ & $2.72\pm0.14$ & - & $0.329$ & $0.041$ & $0.370\pm0.022$ & -\tabularnewline
\hline 
\end{tabular}
\par\end{centering}
\caption{Production cross sections of the 1S and 1P states of the $B_{c}^{+}$
meson in p-Au and Au-Au collisions at RHIC energies. In fourth and
eighth columns, we provide our results using nCTEQ15 LO PDF sets.
In fifth and ninth columns, the estimates of Ref. \cite{Chen:2018obq}
are provided for 1S states. Contributions of the gluon fusion and $q\bar{q}$
annihilation processes to the cross sections in p-Au and Au-Au collisions
at RHIC energy are also provided. The errors due to the uncertainty
in the PDF set are provided only for the total values. \label{tab:cross-au}}
\end{table}

\begin{table}
\begin{centering}
\begin{tabular}{ccccccccc}
\hline 
 & \multicolumn{4}{c|}{$\sigma(pPb\rightarrow B_{c}X)$ ($\mu\textrm{b}$) at $\sqrt{S}=8.16$
TeV} & \multicolumn{4}{c}{$\sigma(PbPb\rightarrow B_{c}X)$ ($\mu\textrm{b}$) at $\sqrt{S}=5.04$
TeV}\tabularnewline
\cline{2-9} \cline{3-9} \cline{4-9} \cline{5-9} \cline{6-9} \cline{7-9} \cline{8-9} \cline{9-9} 
States & nCTEQ15 & EPPS16 & nNNPDF3.0 & Ref & nCTEQ15 & EPPS16 & nNNPDF3.0 & Ref\tabularnewline
\hline 
\hline 
$1^{1}S_{0}$ & $4.19$ & $4.41$ & $4.14$ & $3.29$ & $232$ & $254$ & $225$ & $369$\tabularnewline
$1^{3}S_{1}$ & $10.5$ & $11.0$ & $10.2$ & $8.26$ & $583$ & $638$ & $563$ & $921$\tabularnewline
$1^{1}P_{1}$ & $0.566$ & $0.571$ & $0.569$ & - & $31.4$ & $33.2$ & $30.4$ & -\tabularnewline
$1^{3}P_{0}$ & $0.426$ & $0.446$ & $0.408$ & - & $23.4$ & $25.6$ & $22.7$ & -\tabularnewline
$1^{3}P_{1}$ & $0.681$ & $0.705$ & $0.676$ & - & $37.6$ & $40.4$ & $36.3$ & -\tabularnewline
$1^{3}P_{2}$ & $1.32$ & $1.38$ & $1.22$ & - & $73.1$ & $79.6$ & $70.6$ & -\tabularnewline
\hline 
\end{tabular}
\par\end{centering}
\caption{Comparison of the production cross sections of the 1S and 1P states for p-Pb
and Pb-Pb collisions using different PDF sets. \label{tab:alt_pdf_pb}}
\end{table}

\begin{table}
\begin{centering}
\begin{tabular}{ccccc}
\hline 
 & \multicolumn{2}{c|}{{\footnotesize{}$\sigma(pXe\rightarrow B_{c}X)$ ($\mu\textrm{b}$)
}at $\sqrt{S}=8.16$ TeV} & \multicolumn{2}{c}{{\footnotesize{}$\sigma(XeXe\rightarrow B_{c}X)$ ($\mu\textrm{b}$)
}at $\sqrt{S}=5.44$ TeV}\tabularnewline
\cline{2-5} \cline{3-5} \cline{4-5} \cline{5-5} 
States & nCTEQ15 & nNNPDF3.0 & nCTEQ15 & nNNPDF3.0\tabularnewline
\hline 
\hline 
{\small{}$1^{1}S_{0}$} & $2.81$ & $2.84$ & $104$ & $107$\tabularnewline
{\small{}$1^{3}S_{1}$} & $7.01$ & $7.17$ & $261$ & $267$\tabularnewline
{\small{}$1^{1}P_{1}$} & $0.379$ & $0.383$ & $14.1$ & $14.4$\tabularnewline
{\small{}$1^{3}P_{0}$} & $0.285$ & $0.297$ & $10.5$ & $10.9$\tabularnewline
{\small{}$1^{3}P_{1}$} & $0.456$ & $0.471$ & $16.9$ & $17.4$\tabularnewline
{\small{}$1^{3}P_{2}$} & $0.882$ & $0.905$ & $32.8$ & $33.5$\tabularnewline
\hline 
\end{tabular}
\par\end{centering}
\caption{Comparison of the production cross sections of 1S and 1P states for p-Xe
and Xe-Xe collisions using different PDF sets.\label{tab:alt_pdf_xe}}
\end{table}

\begin{table}
\begin{centering}
\begin{tabular}{ccccr@{\extracolsep{0pt}.}lcc}
\hline 
 & {\footnotesize{}p-Pb} & {\footnotesize{}Pb-Pb} & {\footnotesize{}p-Xe} & \multicolumn{2}{c}{{\footnotesize{}Xe-Xe}} & {\footnotesize{}p-Au} & {\footnotesize{}Au-Au}\tabularnewline
 & {\footnotesize{}($\textrm{\ensuremath{\mu}b}$)} & {\footnotesize{}($\times10\textrm{\ensuremath{\mu}b}$)} & {\footnotesize{}($\textrm{\ensuremath{\mu}b}$)} & \multicolumn{2}{c}{{\footnotesize{}($\times10\textrm{\ensuremath{\mu}b}$)}} & {\footnotesize{}($\textrm{nb}$)} & {\footnotesize{}($\textrm{\ensuremath{\mu}b}$)}\tabularnewline
\hline 
\hline 
{\footnotesize{}$1^{1}S_{0}$} & {\footnotesize{}$4.2_{-0.9-0.7}^{+0.8+0.6}$} & {\footnotesize{}$23_{-5-3}^{+4+3}$} & {\footnotesize{}$2.8_{-0.6-0.4}^{+0.5+0.4}$} & \multicolumn{2}{c}{{\footnotesize{}$10.4_{-2.3-1.6}^{+2.0+1.2}$}} & {\footnotesize{}$8.5_{-2.5-2.2}^{+1.9+1.7}$} & {\footnotesize{}$1.17_{-0.35-0.30}^{+0.26+0.23}$}\tabularnewline
{\footnotesize{}$1^{3}S_{1}$} & {\footnotesize{}$10.5_{-1.3-1.0}^{+1.2+0.5}$} & {\footnotesize{}$58_{-16-8}^{+11+6}$} & {\footnotesize{}$7.0_{-1.5-1.1}^{+1.7+0.9}$} & \multicolumn{2}{c}{{\footnotesize{}$26_{-7-4}^{+5+3}$}} & {\footnotesize{}$23_{-8-5}^{+5+4}$} & {\footnotesize{}$3.0_{-1.0-0.7}^{+0.7+0.6}$}\tabularnewline
{\footnotesize{}$1^{1}P_{1}$} & {\footnotesize{}$0.57_{-0.23-0.09}^{+0.16+0.08}$} & {\footnotesize{}$3.1_{-1.2-0.5}^{+0.9+0.4}$} & {\footnotesize{}$0.38_{-0.15-0.05}^{+0.11+0.05}$} & \multicolumn{2}{c}{{\footnotesize{}$1.4_{-0.6-0.2}^{+0.4+0.2}$}} & {\footnotesize{}$1.2_{-0.6-0.3}^{+0.4+0.2}$} & {\footnotesize{}$0.16_{-0.08-0.04}^{+0.05+0.03}$}\tabularnewline
{\footnotesize{}$1^{3}P_{0}$} & {\footnotesize{}$0.43_{-0.14-0.07}^{+0.10+0.06}$} & {\footnotesize{}$2.3_{-0.8-0.4}^{+0.6+0.3}$} & {\footnotesize{}$0.29_{-0.09-0.05}^{+0.07+0.04}$} & \multicolumn{2}{c}{{\footnotesize{}$1.05_{-0.40-0.17}^{+0.26+0.15}$}} & {\footnotesize{}$1.11_{-0.40-0.23}^{+0.29+0.18}$} & {\footnotesize{}$0.15_{-0.05-0.03}^{+0.04+0.02}$}\tabularnewline
{\footnotesize{}$1^{3}P_{1}$} & {\footnotesize{}$0.68_{-0.26-0.11}^{+0.18+0.09}$} & {\footnotesize{}$3.8_{-1.4-0.6}^{+1.0+0.5}$} & {\footnotesize{}$0.46_{-0.17-0.07}^{+0.12+0.07}$} & \multicolumn{2}{c}{{\footnotesize{}$1.69_{-0.7-0.27}^{+0.4+0.22}$}} & {\footnotesize{}$1.52_{-0.60-0.32}^{+0.40+0.26}$} & {\footnotesize{}$0.21_{-0.09-0.05}^{+0.06+0.03}$}\tabularnewline
{\footnotesize{}$1^{3}P_{2}$} & {\footnotesize{}$1.32_{-0.5-0.20}^{+0.40+0.17}$} & {\footnotesize{}$7.3_{-3.0-1.1}^{+2.0+0.9}$} & {\footnotesize{}$0.88_{-0.40-0.14}^{+0.25+0.11}$} & \multicolumn{2}{c}{{\footnotesize{}$3.3_{-1.3-0.5}^{+0.9+0.4}$}} & {\footnotesize{}$2.7_{-1.3-0.6}^{+0.8+0.5}$} & {\footnotesize{}$0.37_{-0.17-0.07}^{+0.11+0.07}$}\tabularnewline
\hline 
\end{tabular}
\par\end{centering}
\caption{Cross sections of the 1S and 1P states of the $B_{c}^{+}$ meson with uncertainty
due to heavy-quark masses. First and second errors corresponds to
$\Delta m_{c}=\pm0.1$ GeV and $\Delta m_{b}=\pm0.2$ GeV, respectively.
The errors are calculated using nCTEQ15 LO PDF sets. \label{tab:cross-err}}
\end{table}

\subsection{Kinematical distributions}

\begin{figure}[p]
\begin{centering}
\includegraphics[scale=0.58]{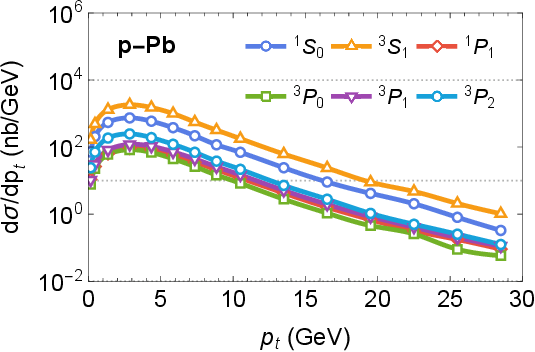}\includegraphics[scale=0.58]{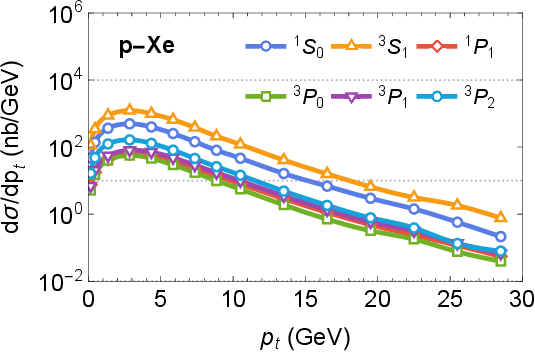}\includegraphics[scale=0.58]{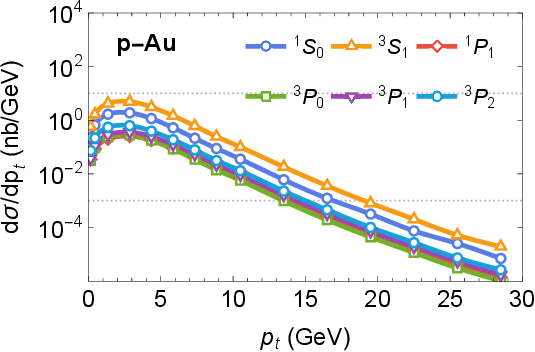}
\par\end{centering}
\caption{$p_{T}$ distributions of the 1S and 1P states of the $B_{c}^{+}$ meson for
p-Pb (left), p-Xe (middle), and p-Au (right) collisions at $\sqrt{S_{pA}}=8.16$,
8.3, 0.2 TeV, respectively.\label{fig:pt-pA}}
\end{figure}

\begin{figure}[p]
\begin{centering}
\includegraphics[scale=0.58]{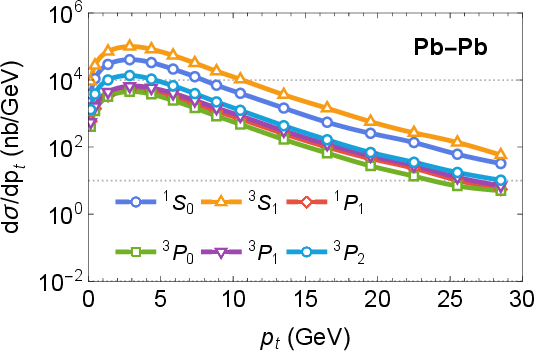}\includegraphics[scale=0.58]{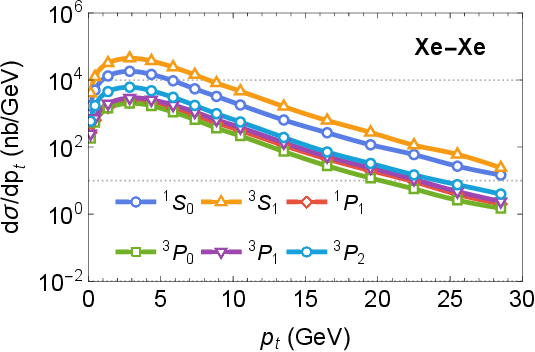}\includegraphics[scale=0.58]{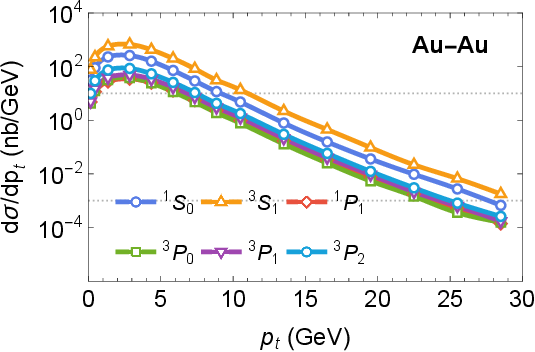}
\par\end{centering}
\caption{$p_{T}$ distributions of the 1S and 1P states of the $B_{c}^{+}$ meson for
Pb-Pb (left), Xe-Xe (middle), and Au-Au (right) collisions at $\sqrt{S_{AA}}=5.02$,
5.44, and 0.2 TeV, respectively.\label{fig:pt-AA}}
\end{figure}

\begin{figure}[p]
\begin{centering}
\includegraphics[scale=0.58]{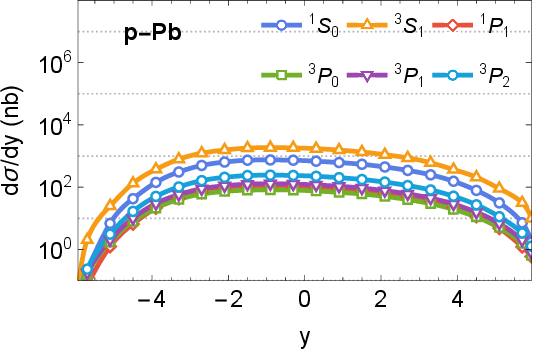}\includegraphics[scale=0.58]{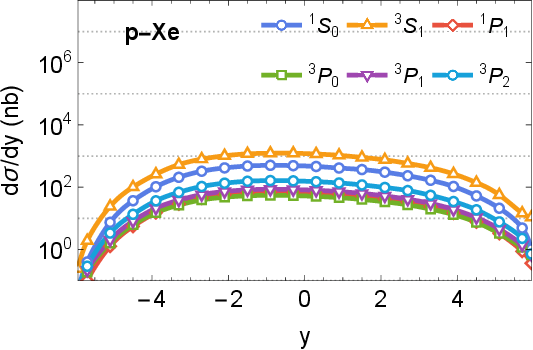}\includegraphics[scale=0.58]{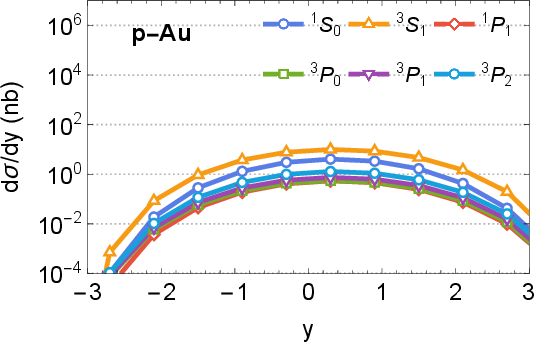}
\par\end{centering}
\caption{$y$ distributions of the 1S and 1P states of the $B_{c}^{+}$ meson for p-Pb
(left), p-Xe (middle), and p-Au (right) collisions at $\sqrt{S_{pA}}=8.16$,
8.3, 0.2 TeV, respectively.\label{fig:y-pA}}
\end{figure}

\begin{figure}[p]
\begin{centering}
\includegraphics[scale=0.58]{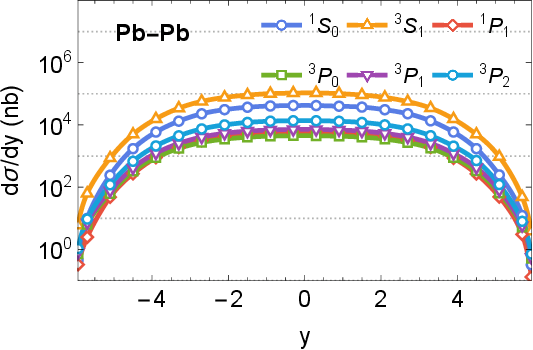}\includegraphics[scale=0.58]{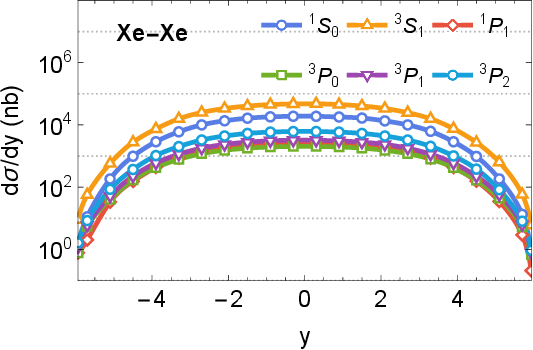}\includegraphics[scale=0.58]{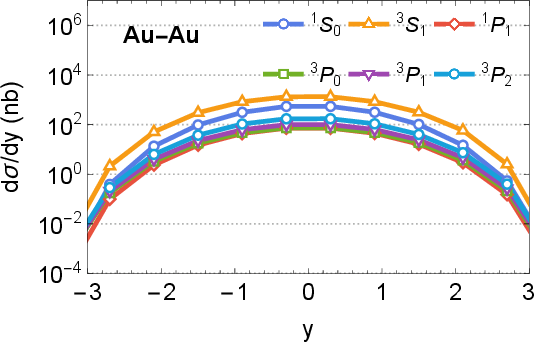}
\par\end{centering}
\caption{$y$ distributions of the 1S and 1P states of the $B_{c}^{+}$ meson for Pb-Pb
(left), Xe-Xe (middle), and Au-Au (right) collisions at $\sqrt{S_{AA}}=5.02$,
5.44, and 0.2 TeV, respectively.\label{fig:y-AA}}
\end{figure}

\begin{figure}[p]
\begin{centering}
\includegraphics[scale=0.58]{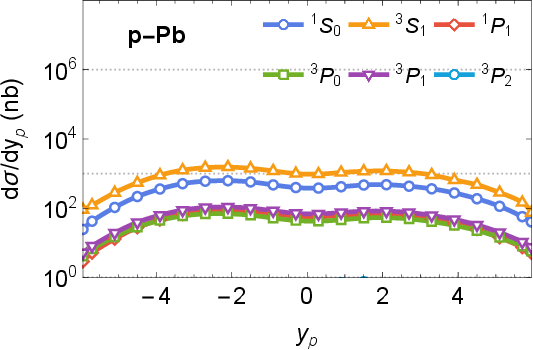}\includegraphics[scale=0.58]{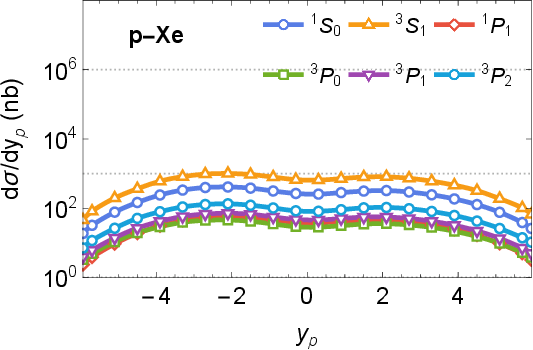}\includegraphics[scale=0.58]{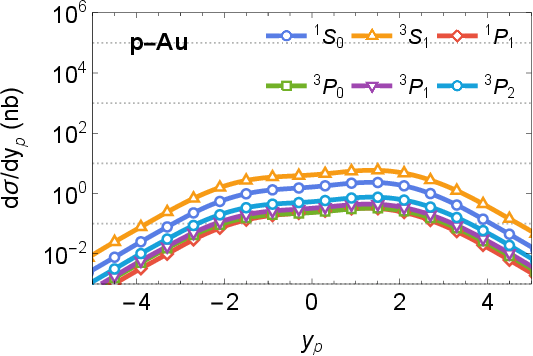}
\par\end{centering}
\caption{$y_{P}$ distributions of the 1S and 1P states of the $B_{c}^{+}$ meson for
p-Pb (left), p-Xe (middle), and p-Au (right) collisions at $\sqrt{S_{pA}}=8.16$,
8.3, 0.2 TeV, respectively.\label{fig:yp-pA}}
\end{figure}

\begin{figure}[p]
\begin{centering}
\includegraphics[scale=0.58]{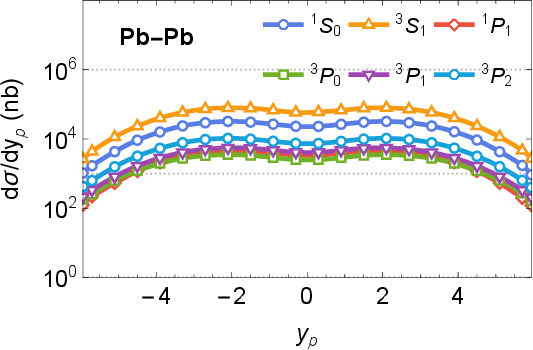}\includegraphics[scale=0.58]{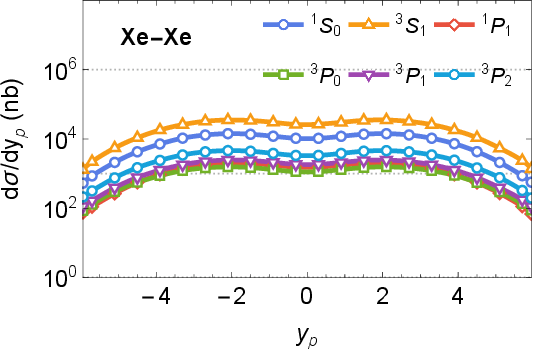}\includegraphics[scale=0.58]{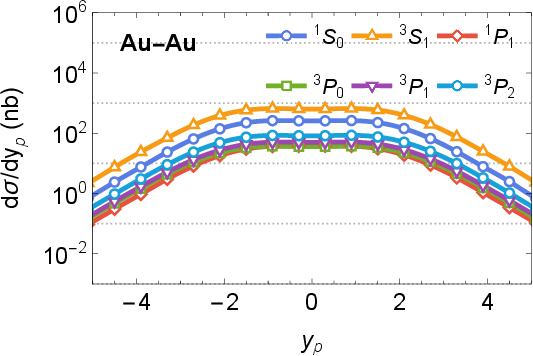}
\par\end{centering}
\caption{$y_{P}$ distributions of the 1S and 1P states of the $B_{c}^{+}$ meson for
Pb-Pb (left), Xe-Xe (middle), and Au-Au (right) collisions at $\sqrt{S_{AA}}=5.02$,
5.44, and 0.2 TeV, respectively.\label{fig:yp-AA}}
\end{figure}

In Figs. \ref{fig:pt-pA} and \ref{fig:pt-AA}, we show the plots
of $p_{T}$ distributions of the 1S and 1P states of the $B_{c}^{+}$ meson
in pA and AA collisions, respectively, for Pb, Xe, and Au nuclei. All
distributions have a similar shape: increasing rapidly, exhibiting a
maxima at about $p_{T}=2.5$ GeV and then decreasing slowly. Beyond
$p_{T}=10$ GeV, the differential cross section is reduced by one
order of magnitude and and beyond $20$ GeV, it is reduced by two
orders of magnitude. In Figs. \ref{fig:y-pA} and \ref{fig:y-AA},
we show the plots of $y$ (rapidity) distributions of the 1S and 1P states
of the $B_{c}^{+}$ meson. The distributions of pA collisions are asymmetric
due to the uneven energy of colliding particles. The asymmetry for p-Pb
and p-Xe collisions is relatively mild as compared to p-Au collision.
We also observe that the plateau in mid rapidity for pA and AA collisions
at LHC energies is very broad, extending over $-2<y<2$, whereas for
pAu and Au-Au collisions at RHIC energy, the distributions are round-shaped. In Figs. \ref{fig:yp-pA} and \ref{fig:yp-AA}, we show the
plots of $y_{p}$ (pseudo rapidity) distributions of the 1S and 1P states
of the $B_{c}^{+}$ meson. The distributions of pA collisions are once again
asymmetric due to the uneven energy of colliding particles, and plateau
in case of collision at LHC energies is relatively broad.

\subsection{Effect of kinematical cuts}

\begin{figure}[p]
\begin{centering}
\includegraphics[scale=0.58]{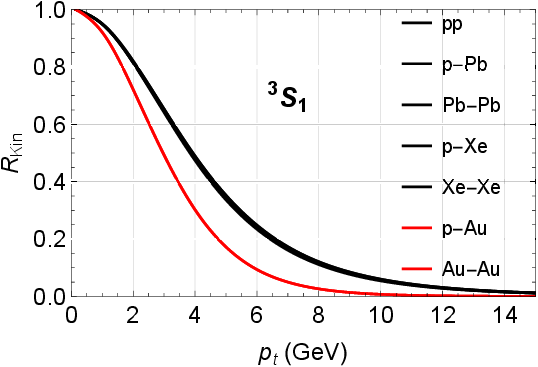}\includegraphics[scale=0.58]{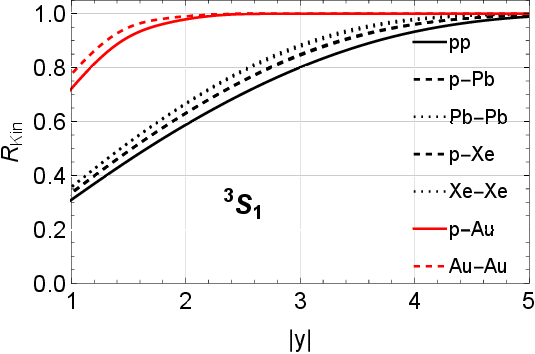}\includegraphics[scale=0.58]{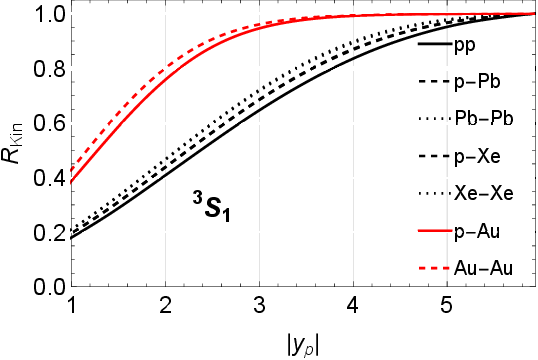}
\par\end{centering}
\caption{Variation of $R_{kin}$ with the values of cuts on $p_{T}$ (left), $\left|y\right|$
(middle), and $\left|y_{p}\right|$ (right). Cuts in the case $\left|y\right|$
and $\left|y_{p}\right|$ set upper limit ($x<x_{0})$.\label{fig:cut-pt-3s1}}
\end{figure}

\begin{figure}[p]
\begin{centering}
\includegraphics[scale=0.7]{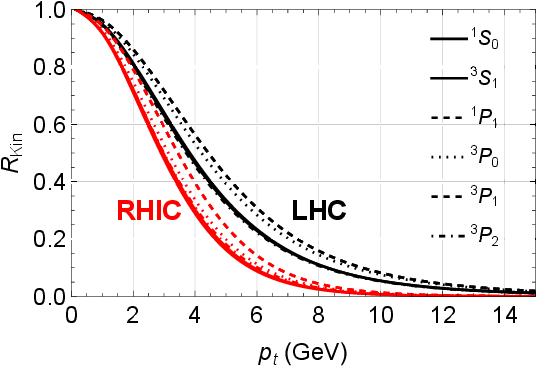}
\includegraphics[scale=0.65]{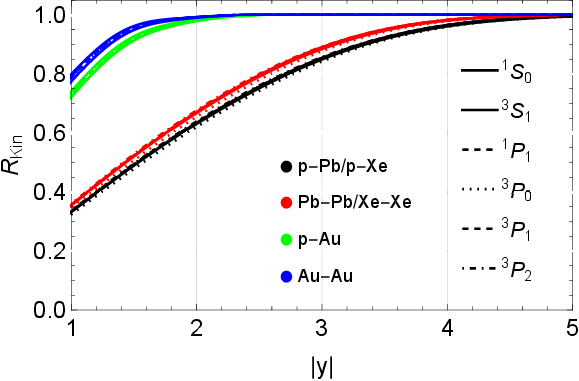}
\par\end{centering}
\caption{Left: Variation of $R_{kin}$ with the values of cuts on $p_{T}$ cut for different states
of the $B_{c}$ meson. Collision mode at LHC and RHC energies have no
impact on $R_{Kin}$ as implied by the plots of Fig. \ref{fig:cut-pt-3s1}.
Right: Variation of $R_{Kin}$ with $\left|y\right|$ cut for different
states of the $B_{c}$ meson. We use different color codes to separate
collision modes and different line formates to separate states. \label{fig:cut-pt-states}}
\end{figure}

The values of total cross sections reported in the table \ref{tab:cross-err}
are obtained without any kinematical cuts on $p_{T}$ and $y$ (or $y_{p}$)
of the $B_{c}$ meson. A particle with a small value of $p_{T}$ or large
value of rapidity moves very close to beam axis and due to cylindrical
geometry of a high energy particle detector, it is impossible to detect
it. Therefore, we typically measure cross sections that are subjected to kinematical
cuts on $p_{T}$ and rapidity. Considering these limitations, we also
study the effect of kinematical cuts on the production cross sections
of $1S$ and $1P$ states of the $B_{c}$ meson. To quantify the effect
of the cuts in a more general way, we define following ratio:
\begin{equation}
R_{kin}(x_{0})=\frac{\left(\sigma_{pA/AA}\right)_{x>x_{0}}}{\left(\sigma_{pA/AA}\right)_{\textrm{no cut}}},\label{eq:ratio-cut}
\end{equation}
where $x=p_{T}$, $\left|y\right|$, $\left|y_{p}\right|$. In the
denominator, we have total cross section without a kinematical cut, and
in numerator, we have it with the kinematical cut $x>x_{0}$. It is also
noted that for $\left|y\right|$ and $\left|y_{p}\right|$, the cut
sets upper limit, so it is expressed by $x<x_{0}$. First, we chose
$^{3}S_{1}$ state of $B_{c}^{+}$ as a reference and determine how $R_{kin}$
varies with $x_{0}$ in the case of pA and AA collisions, taking A =
Pb, Xe, and Au. In Fig. \ref{fig:cut-pt-3s1}, we show the plots of the
variation in $R_{kin}(x_{0})$ with a cut on $p_{T},$ $\left|y\right|$,
and $\left|y_{p}\right|$ respectively. The left plot in the figure shows
that the effect of the cut on $p_{T}$ is nearly same for all collisions modes at LHC
energies. Similarly, the effect is same for the collision
modes (p-Au and Au-Au) at RHIC energies, but different from the collision
modes at LHC energies. However, the middle and right plots given in
the figure show that the impact of cuts on $y$ and $y_{p}$ is influenced
not only by the collision energy but also by the collision mode. The middle plot of Fig. \ref{fig:cut-pt-3s1} for p-Au and Au-Au
modes show that the values of cross sections saturate for cut on $\left|y\right|>2$,
whereas no saturation effect is observed for collision modes at LHC
energies. Similarly, the right plot of the figure for p-Au and Au-Au
modes show saturation effect for cut on $\left|y_{p}\right|>3$, but
no saturation effect for collision modes at LHC energies. 
In Fig. \ref{fig:cut-pt-states} (left), we show the variation of
$R_{kin}$ with value of cut on $p_{T}$ for different states of the $B_{c}$
meson. In this case, we refrain from distinguishing between collision modes
because the profile of $R_{kin}$ remains unaffected by this distinction, as
indicated by the left plot of Fig. \ref{fig:cut-pt-3s1}. Since the
profile does depend on the energy scale, so we use red and black
colors to distinguish LHC and RHIC results. The plot also
shows that the effect of kinematical cut on $p_{T}$ slightly varies with states of the $B_{c}$ meson. In Fig. \ref{fig:cut-pt-states}
(right), we show the variation of $R_{kin}$ with the value of cut
on $\left|y\right|$ for different states of $B_{c}$ meson. In this
case, we do make a distinction between collision modes by using black
(p-Pb, p-Xe), red (Pb-Pb, Xe-Xe), green (p-Au), and blue (Au-Au)
colors. The plots show that the effect due to variation in the state
of the meson is negligible. 

\subsection{Nuclear Modification Factors}

\begin{figure}
\begin{centering}
\includegraphics[scale=0.58]{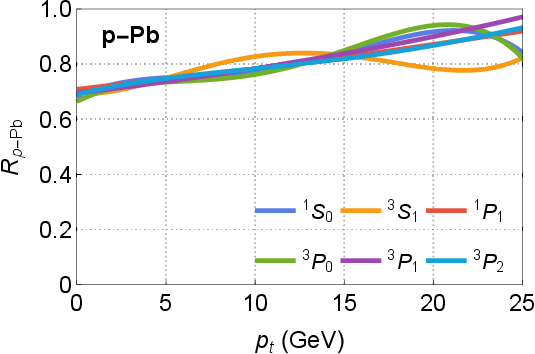}\includegraphics[scale=0.58]{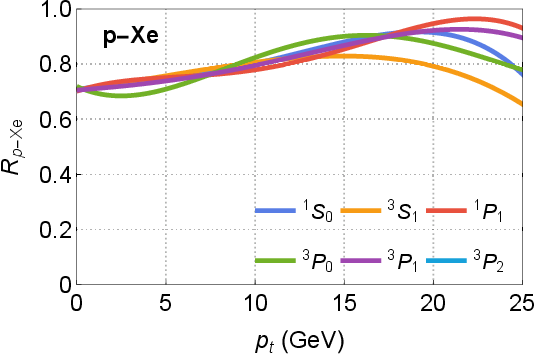}\includegraphics[scale=0.58]{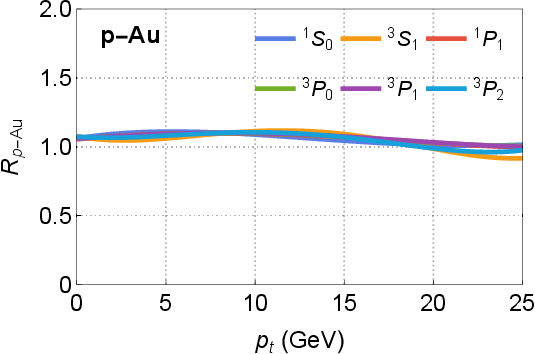}
\par\end{centering}
\caption{$p_{T}$ distributions of the nuclear modification factors $R_{pA}$ of
the 1S and 1P states of the $B_{c}$ meson for p-Pb (left), p-Xe (middle), and
p-Au (right) collisions. \label{fig:rp-pt-pA}}
\end{figure}

\begin{figure}
\begin{centering}
\includegraphics[scale=0.58]{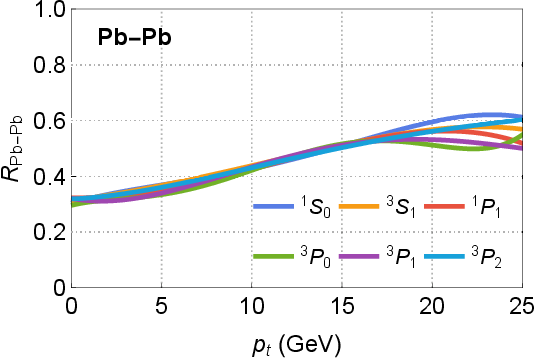}\includegraphics[scale=0.58]{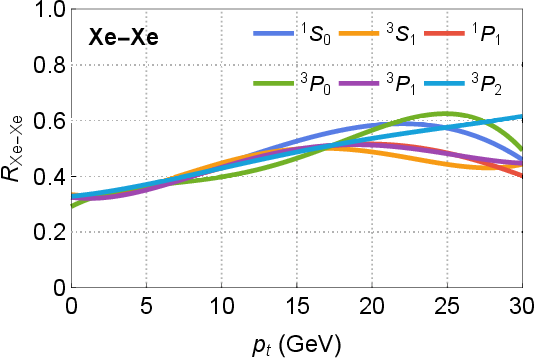}\includegraphics[scale=0.58]{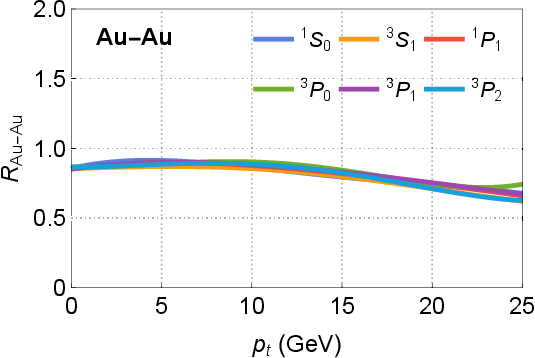}
\par\end{centering}
\caption{$p_{T}$ distributions of the nuclear modification factors $R_{AA}$ of
the 1S and 1P states of $B_{c}$ meson for Pb-Pb (left), Xe-Xe (middle),
and Au-Au (right) collisions. \label{fig:rp-pt-AA}}
\end{figure}

\begin{figure}
\begin{centering}
\includegraphics[scale=0.58]{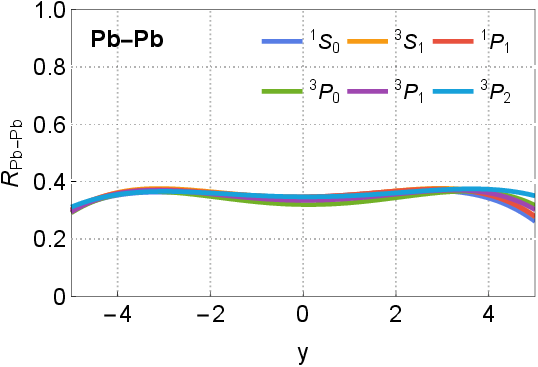}\includegraphics[scale=0.58]{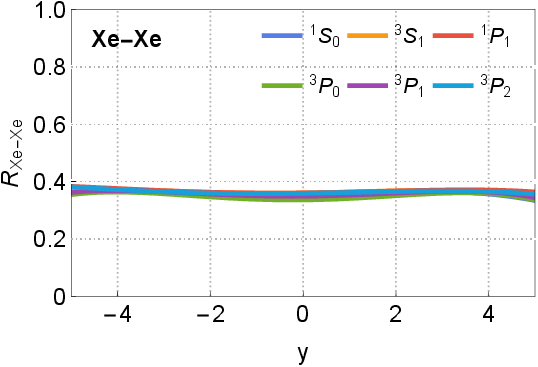}\includegraphics[scale=0.58]{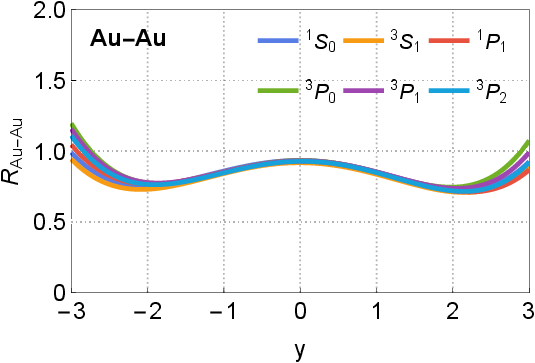}
\par\end{centering}
\caption{$y$ distributions of the nuclear modification factors $R_{AA}$ of the 1S and
1P states of the $B_{c}$ meson for Pb-Pb (left), Xe-Xe (middle), and Au-Au
(right) collisions.\label{fig:rp-y-AA}}
\end{figure}

Starting from the basic formula nuclear modification factor for case
of minimum bias and expressed in terms of partial cross sections
\begin{equation}
R_{AB}^{(h)}=\frac{\left(d\sigma_{AB}^{(h)}\right)^{(th)}}{N_{A}N_{B}\left(d\sigma_{pp}^{(h)}\right)},
\end{equation}
we can define $p_{T}$ and rapidity dependent nuclear modification
factors as following
\begin{eqnarray}
R_{AB}^{(h)}(p_{T}) & = & \frac{\left(d\sigma_{AB}^{(h)}/dp_{T}\right)^{(th)}}{N_{A}N_{B}\left(d\sigma_{pp}^{(h)}/dp_{T}\right)},\\
R_{AB}^{(h)}(y) & = & \frac{\left(d\sigma_{AB}^{(h)}/dy\right)^{(th)}}{N_{A}N_{B}\left(d\sigma_{pp}^{(h)}/dy\right)},
\end{eqnarray}
where for pA collision $N_{B}=1$ and for AA collision $N_{A}=N_{B}$.
We use these definitions to calculate $p_{T}$ and rapidity distributions
for pA and AA collisions for Pb and Xe nuclei at LHC energies and
for Au at RHIC energy. In Fig. \ref{fig:rp-pt-pA}, we show the plots
of $p_{T}$ distributions of nuclear modification factor $R_{pA}$
of S and P states of $B_{c}^{+}$ meson for p-Pb, p-Xe, and p-Au collisions
respectively at the cm of energies given in table \ref{tab:cm-energies}.
In each case, we are required to re-calculate $p_{T}$ distributions
of the $B_{c}^{+}$ states in pp collision at $\sqrt{S}=8.16$, $8.30$,
and $0.2$ TeV for p-Pb, p-Xe, and p-Au collisions respectively. Additionally,
in calculating nuclear modification factor for p-Au collision, we
also include the contribution of $q\bar{q}$ annihilation processes
to $B_{c}$ production cross section $(d\sigma_{pp}^{(h)})$ in pp
collison appearing in the denominator as it is already done for corresponding
cross section $(d\sigma_{pAu}^{(h)})$ in p-Au collision appearing
in the numerator. In Fig. \ref{fig:rp-pt-AA}, we show the plots of
$p_{T}$ distributions of nuclear modification factor $R_{AA}$ of
S and P states of $B_{c}^{+}$ meson for Pb-Pb, Xe-Xe, and Au-Au collisions
respectively at the cm of energies given in table \ref{tab:cm-energies}.
In each case, we are required to re-calculate $p_{T}$ distributions
of the $B_{c}^{+}$ states in pp collision at $\sqrt{S}=5.04$, $5.44$,
and $0.2$ TeV for Pb-Pb, Xe-Xe, and Au-Au collisions respectively.
Again, in calculating nuclear modification factor for Au-Au collision,
we include the contribution of $q\bar{q}$ annihilation processes.
In Fig. \ref{fig:rp-y-AA}, we show the plots of rapidity distributions
of nuclear modification factors $R_{AA}$ of S and P states for $A=$
Pb, Xe, and Au at the cm of energies given in table \ref{tab:cm-energies}. 

\section{Conclusion}

In this work, we have obtained several new results regarding the production
of cross sections for various states of the $B_{c}$ meson in heavy-ion
collisions. Previous estimates of these cross sections were available
for $1^{1}S_{0}$ and $1^{3}S_{1}$ states for the p-Pb, Pb-Pb collisions
at LHC energies and for the p-Au and Au-Au collisions at RHIC energy.
We have extended the calculations to include the spin singlet state $1^{1}P_{1}$
and the spin triplet states $1^{3}P_{0}$, $1^{3}P_{1}$, and $1^{3}P_{2}$.
Additionally, we have included p-Xe and Xe-Xe collision modes at LHC
energies, along with p-Pb, Pb-Pb, p-Au, and Au-Au modes. Apart from
calculating $p_{T}$, rapidity, and pseudo rapidity distributions
of cross sections, we have also determined the nuclear modification factors
$R_{pA}$ and $R_{AA}$ for Pb, Xe, and Au nuclei. In this section, we schematically summarize these
results. 
\begin{itemize}

\item Comparison with previous estimates: Currently, the estimates of cross sections for the $B_{c}$ mesons
are available at LO order and only for $1S$ and $1P$ states in pp
collision. Recently, the cross sections for 1S states are provided for
heavy-ion collisions, including p-Pb, Pb-Pb, p-Au, Au-Au modes. To validate
the accuracy of our computations, we first reproduced these results.
Comparison of our results with those given in Ref. \citep{Chang:2005bf}
for the production cross sections of 1S and 1P states in pp collision shows a discrepancy
of less than 5\% after matching the PDF set and input parameters, as indicated
in Fig. \ref{fig:pp_comp}. For 1S states in p-Pb and Pb-Pb modes, the
average difference between our results and that of Ref. \citep{Chen:2018obq}
is about 30\% which reduce to 20\% for p-Au and Au-Au modes. It was
not possible to identify the cause of this discrepancy because the authors
do not provide any information about the choice of $\alpha_{s}$.
Nevertheless, the difference is relatively small as compared to the uncertainties
due to errors in the input parameters and PDF sets. 

\item New Results of the production cross sections in pA and AA collisions: Our
results of the production cross sections for p-Pb, Pb-Pb, p-Xe, Xe-Xe,
p-Au, and Au-Au collision modes are provided in tables \ref{tab:cross-pb},
\ref{tab:cross-xe}, and \ref{tab:cross-au}. The results show that
the production cross sections of $1^{1}P_{1}$, $1^{3}P_{0}$, $1^{3}P_{1}$,
and $1^{3}P_{2}$ states are almost 5\%, 5\%, 7\%, and 13\% of the cross
section of $1^{3}S_{1}$ state, respectively, irrespective of the collision
mode and the collision energy. Whereas, the cross section $^{1}S_{0}$
state is almost 40\% of $1^{3}S_{1}$ state. 

\item $p_{T}$ distributions of the cross sections: In Fig. \ref{fig:pt-pA}
and \ref{fig:pt-AA}, we show the plots of the $p_{T}$ distributions
of 1S and 1P states of the $B_{c}^{+}$ meson for pA and AA collisions.
These distributions have similar shape: increasing rapidly, exhibiting
a maxima at about $p_{T}=2.5$ GeV and then decreasing slowly. Beyond
$p_{T}=10$ GeV, the differential cross section is reduced by one
order of magnitude and and beyond $p_{T}=20$ GeV it is reduced by
two orders of magnitude.

\item Rapidity $(y)$ distributions of the corss sections: In Fig. \ref{fig:y-AA}
and \ref{fig:y-pA}, we show the plots of $y$ (rapidity) distributions
of the 1S and 1P states of the $B_{c}^{+}$ meson in pA and AA collisions, respectively.
The distribution for pA collisions are asymmetric due to the uneven energy
of colliding particles. The asymmetry for p-Pb and p-Xe collisions
is relatively mild as compared to p-Au collisions. We also observe
that the plateau in mid-rapidity for pA and AA collisions at LHC energies
is very broad, extending over $-2<y<2$, whereas for collisions at
RHIC energy, distribution are more round shaped.

\item Effect of kinematical cuts: To quantify the effect of kinematical
cuts on production cross sections, we have introduced a ratio $R_{kin}$
given in Eq. \ref{eq:ratio-cut}. We calculate it for different
values of cuts on $p_{T}$ or $\left|y\right|$. The resulting profile
of the ratio provides a convenient method to estimate the
cross sections for any given value of applied cut. In Fig. \ref{fig:cut-pt-states},
we show the variation of $R_{kin}$ with the cut on $p_{T}$
and $\left|y\right|$, respectively, for different states of the $B_{c}$
meson and for different collision modes. We use different color codes
and line formates to distinguish collision modes and states, respectively.
The plots in the Fig. \ref{fig:cut-pt-states} (left) show that the
effect of kinematical cut on $p_{T}$ varies slightly for different
states states of the $B_{c}$ meson. It also depends on the collision
energy but not on specific mode of collision at a given collision energy.
Whereas the plots in the Fig. \ref{fig:cut-pt-states} (right) show
that the effect of cut on rapidity does not depend on the states.
Dependence on collision mode (p-A or A-A) is mild but significant
on collision energy. For p-Au and Au-Au modes, the plots show that
the values of cross sections saturate for cuts on $\left|y\right|>2$,
however no saturation effect is observed for collision modes at LHC
energies.

\item Uncertainties due heavy quark masses: To study the effect of errors
in masses of charm and bottom quarks, we calculate the cross sections
for three values of $m_{c}$ (1.4, $1.5$, and $1.6$ GeV) while taking
$m_{b}=4.9$ GeV. Similarly we calculate cross sections for three
values of $m_{b}$ (4.7, $4.9$, and $5.1$ GeV) while taking $m_{c}=1.5$
GeV. In table \ref{tab:cross-err}, we combine all these results and
report the values of cross sections with uncertainties due to heavy
quark masses. In all cases, the uncertainty due to error in the b-quark
mass is almost half that of the c-quark mass. The results show that the
errors due $m_{c}$ and $m_{b}$ masses are about 25\% and 13\%, respectively,
for p-Pb, Pb-Pb, p-Xe, and Xe-Xe modes. And for p-Au and Au-Au modes,
the errors are about 30\% and 20\%, respectively. 

\item Nuclear Modification factors: In Fig. \ref{fig:rp-pt-pA}, we show
the plots of the  $p_{T}$ distributions of the nuclear modification factor
$R_{pA}$ of the S and P states of the $B_{c}^{+}$ meson for p-Pb, p-Xe,
and p-Au collisions, respectively. The profiles of $R_{pA}$ are similar for
p-Pb and p-Xe modes. In both cases, $R_{pA}$ exhibits almost linear
increase with $p_{T}$, ranging from approximately 0.7 to 1 for $p_{T}=$
0 to 25 GeV. However, for p-Au mode at RHIC energy, $R_{pA}\approx1$.
In Fig. \ref{fig:rp-pt-AA}, we show the plots of $p_{T}$ distributions
of the nuclear modification factor $R_{AA}$ of the 1S and 1P states of the $B_{c}^{+}$
meson for Pb-Pb, Xe-Xe, and Au-Au collisions respectively. The profiles of
$R_{AA}$ are similar for Pb-Pb and Xe-Xe modes. In both cases, $R_{AA}$
exhibits almost linear variation with $p_{T}$, ranging from approximately
0.35 to 0.6, almost half of $R_{pA}$. However, for Au-Au mode, $R_{AA}\approx0.9$
up to $p_{T}=10$ GeV, and it gradually decreases to 0.7 at
$p_{T}=25$ GeV. In Fig. \ref{fig:rp-y-AA}, we show the plots of
the rapidity distributions of the nuclear modification factor $R_{AA}$ for
1S and 1P states in Pb-Pb, Xe-Xe, and Au-Au collisions, respectively,
at the relevant c.m. energies. The profiles of $R_{AA}$ are similar
for Pb-Pb and Xe-Xe modes. In both cases, $R_{AA}$ exhibits almost
a constant value of 0.4. However, for Au-Au mode, it is almost 0.9. We do
not present the plots of the rapidity distributions of the nuclear modification
factor $R_{pA}$. These distributions are not useful, because the
scaled rapidity distribution (the denominator of the $R_{pA}$) is
calculated in the c.m. frame, while the distribution with nuclear PDF
(the numerator of the $R_{pA}$) is not in the c.m. frame. Since rapidity
is not boost invariant, the resulting profiles of $R_{pA}$ do not
truly quantify the modification in cross sections due to nuclear
shadowing.
\end{itemize}

This study is focused on LO estimates of the production cross sections
and nuclear modification factors of the 1S and 1P states of the $B_{c}$ meson
at LHC and RHIC energies. The results exhibit significant dependance
on the masses of heavy-quarks and the factorization scale. We expect that NLO
results would help addressing these issues. Nevertheless, our results
provide a baseline for upcoming experimental studies of $B_{c}$ mesons
in pp, pA, and AA collisions in RUN III of LHC and could help in developing
optimal experimental search strategies for identification of its new states. Moreover,
with RUN III's higher luminosity, LHC would be able to measure production
cross sections and distributions, providing more stringent constraints
to test the validity of NRQCD approach for charm-beauty mesons.

\bibliographystyle{unsrt}
\bibliography{paper1v3}

\end{document}